\documentclass[useAMS,usenatbib]{mn2e}
\usepackage{graphicx}
\usepackage{subfigure}
\usepackage{amsmath}
\usepackage{lmodern}
\usepackage{url}
\usepackage{breqn}

\usepackage{acronym}

\acrodef{AO}[AO]{adaptive optics}
\acrodef{SNR}[SNR]{signal to noise ratio}

\graphicspath{{./figs/}}

\title[Improved shift estimates on extended images]{Improved shift estimates on extended Shack-Hartmann wave-front sensor images}

\author[M. J. Townson, A. Kellerer, and C. D. Saunter]{M. J. Townson$^{1}$, A. Kellerer$^{2}$, and C. D. Saunter$^{1}$\thanks{E-mail: christopher.saunter@durham.ac.uk}\\
$^{1}$Department of Physics, Durham University, South Road, DH1 3LE, UK\\
$^{2}$University of Cambridge, Cavendish Laboratory, JJ Thomson Av., Cambridge, CB30HE, UK}

\begin{document}

\date{in original form 2015 February 16}
\pagerange{\pageref{firstpage}--\pageref{lastpage}} \pubyear{2015}

\maketitle

\label{firstpage}

\begin{abstract}
An important factor which affects performance of solar \ac{AO} systems is the accuracy of tracking an extended object in the wave-front sensor. The accuracy of a center of mass approach to image shift measurement depends on the parameters applied in thresholding the recorded image, however there exists no analytical prediction for these parameters for extended objects. Motivated by this we present a new method for exploring the parameter space of image shift measurement algorithms, and apply this to optimise the parameters of the algorithm. Using a thresholded, windowed center of mass, we are able to improve centroid accuracy compared to the typical parabolic fitting approach by a factor of $3 \times$ in a signal to noise regime typical for solar \ac{AO}. Exploration of the parameters occurs after initial image cross-correlation with a reference image, so does not require regeneration of correlation images. The results presented employ methods which can be used in real-time to estimate the error on centroids, allowing the system to use real data to optimise parameters, without needing to enter a separate calibration mode. This method can also be applied outside of solar \ac{AO} to any field which requires the tracking of an extended object.
\end{abstract}


\section{Introduction}
\label{sec:introduction}
Tracking extended objects from image sequences in the presence of noise is required in many different fields. Within astronomy it is used in \ac{AO}, both for granular images of the sun during the day \citep{Scharmer2003,Rimmelea1998,Michau1993}, and elongated laser guide stars at night \citep{Thomas2008}. Image shift measurement is also used in other fields, for tracking biological samples \citep{Hand2009}, and video motion tracking applications.

\begin{figure}
    \begin{center}
        \includegraphics{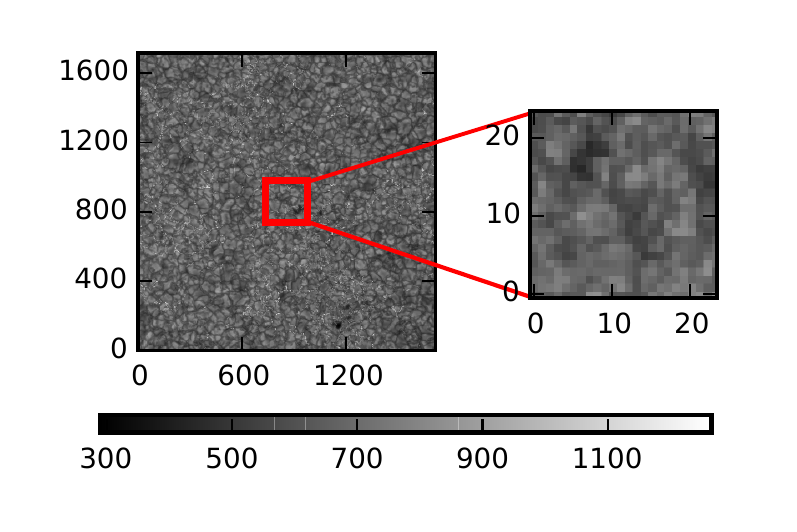}
    \end{center}
    \caption{Image of solar granulation used as the input image in the simulations. The full image is $75 \times 75$ arc-seconds. Small regions of the image are taken and then shifted with respect to each-other to artificially generate shifts similar to the effect of turbulence of the atmosphere. One such region is shown to the right of the full image. It has been re-sampled to the resolution used in the simulations, $0.4 \mathrm{arc-seconds}/\mathrm{pixel}$. Data obtained from the Swedish Solar Telescope On-line Gallery \citep{sst_gallery}.}
    \label{fig:granules}
\end{figure}

In solar \ac{AO}, Shack-Hartmann wave-front sensors \citep{shack1971} with large fields of view are typically employed. The cameras used in these sensors have large full well depths, as the instruments are photon-noise limited. In this work, we investigate tracking extended objects using data acquired from the Swedish Solar Telescope on-line gallery \citep{Scharmer1999a} as our wave-front sensor images (Fig.~\ref{fig:granules}), which have an rms contrast of $10\%$. We assume a photon-noise limited camera with a signal defined as the peak intensity above the background, and a noise level defined by the photon-noise. For camera pixels with a typical full well depth of 40000 electrons; the signal would be 4000 electrons, with photon-noise from 40000 electrons, giving a \ac{SNR} of 20.

Shift measurements of extended objects in solar \ac{AO} are calculated in a two-step process \citep{Michau2006}. Initially an integer shift measurement is performed by locating the peak of a cross-correlation of the image with a reference image \citep{Miura2009}. Secondly the sub-pixel shift is estimated. The determination of the peak location to sub-pixel accuracy limits the accuracy to which the shift measurement can be performed. 

We concern ourselves with how to best estimate the peak location to a sub-pixel accuracy for an arbitrarily shaped correlation function derived from cross correlating the object with a reference image. For point sources and the resultant Airy functions there are analytical methods to determine optimal parameters for peak location at a given \ac{SNR} \citep{Pan2008}. However, no such analytical treatment exists for images of arbitrary content, such as the results of a correlating wave-front sensor. Motivated by this we developed a method to optimise the parameters for a windowed, thresholded, center of mass measurement for a given \ac{SNR}. We compare this technique with an analytic 2D parabolic fit to the central $3 \times 3$ pixel region around the correlation peak, as described in \citet{Lofdahl2010}. This method was chosen as a comparison as its performance is similar to the 2D quadratic interpolation method, and significantly better than the 1D techniques \citep{Lofdahl2010} and Gaussian fitting algorithms \citep{Waldmann2007}.

\section{Correlation Image Generation}
\label{sec:simulation}
Simulations for this paper were run on images containing solar granulation of a size, field of view and contrast typical for solar \ac{AO}, taken from the large image shown in Fig.~\ref{fig:granules} \citep{Scharmer1999a}. The image was shifted and binned in order to generate images containing sub-pixel shifts using the Python language, and numpy routines \citep{Vanderwalt2011}. The solar granulation case used here is an example to demonstrate the use of the centroiding technique, though in general it should be applicable to any extended image.

Most of the computational load in centroiding extended objects lies in cross-correlating images. By varying parameters applied to centroiding the correlation images, the same correlation image can be used.

Regions of $240 \times 240$ pixels were taken from Fig.~\ref{fig:granules}, corresponding to a $9.6$ arc-second field of view. Integer shifts were performed on the full resolution image, with a Gaussian distribution of mean 0 and standard deviation of 1 pixel, then the resultant images were binned by a factor of 10 and had shot noise applied, creating typical sub-aperture images of $24 \times 24$ pixels, making fully described shifted images down to 0.1 pixels. These values were chosen to be indicative of residuals in a closed loop \ac{AO} system. The images, along with the known applied shifts were used to compare the windowed 2D parabolic fit \citep{Lofdahl2010} and the windowed, thresholded center of mass methods.

\section{Peak Location on a Correlation Image}
\label{sec:centroiding}

\subsection{Windowed Parabolic Fit}
\label{sec:parabolic}
A small $3 \times 3$ region around the peak of the correlation image can be fitted by a 2D parabola, as described in \citet{Lofdahl2010}. The parabola takes the form:

\begin{equation}
    f(x, y) = a_1 + a_2x + a_3x^2 + a_4y + a_5y^2 + a_6xy,
\end{equation}
where the location of the minima, in $x$ and $y$ respectively are given analytically by:
\begin{align}
    x_{\text{min}} &= i_{\text{min}} + (2a_2a_5 - a_4a_6)/(a^2_6 - 4a_3a_5) \\
    y_{\text{min}} &= j_{\text{min}} + (2a_3a_4 - a_2a_6)/(a^2_6 - 4a_3a_5).
\end{align}
where $i_{\text{min}}$ and $j_{\text{min}}$ are the integer positions of the peak of the correlation in $x$ and $y$ respectively, and the solution to a least squares fit can be found analytically:
\begin{equation}
        \begin{array}{l l l l l}
        a_2     &= \left( \left< s_{1,j} \right>_j - \left< s_{-1,j} \right> _j \right) /2 \\
        a_3     &= \left( \left< s_{1,j} \right> _j - 2\left< s_{-1,j} \right> _j + \left< s_{-1,j} \right> _j\right)/2 \\
        a_4     &= \left( \left< s_{i,1} \right> _i - \left< s_{i,-1} \right> _i \right)/2 \\
        a_5     &= \left( \left< s_{i,1} \right> _i - 2 \left< s_{i,0} \right> _i + \left< s_{i, -1} \right> _i \right) /2 \\
        a_6 &= \left(s_{1,1} - s_{-1,1} - s_{1,-1} + s_{-1,-1} \right) /4
        \end{array} .
    \label{eqn:2d_quad}
\end{equation}
where $s$ describes the $3 \times 3$ windowed region around the correlation peak, $s_{i,j}$ describes the $i^{\text{th}}$ and $j^{\text{th}}$ element of $s$, and $i$,$j$ can take values from $-1$ to $1$ around the center of the peak (located at $s_{0,0}$).

\begin{figure}
    \centering
        \subfigure[]{\label{fig:aliasing_a}
                   \includegraphics{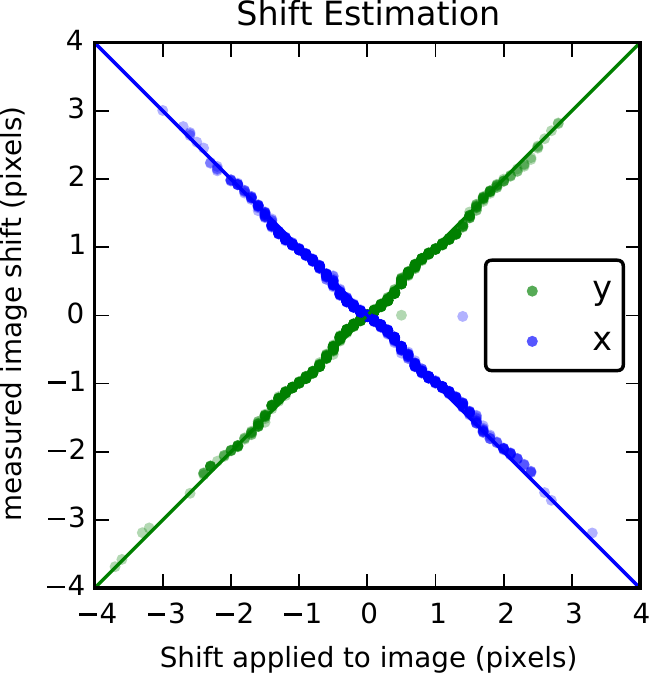}
                   }
        \\
        \subfigure[]{\label{fig:aliasing_b}
                   \includegraphics{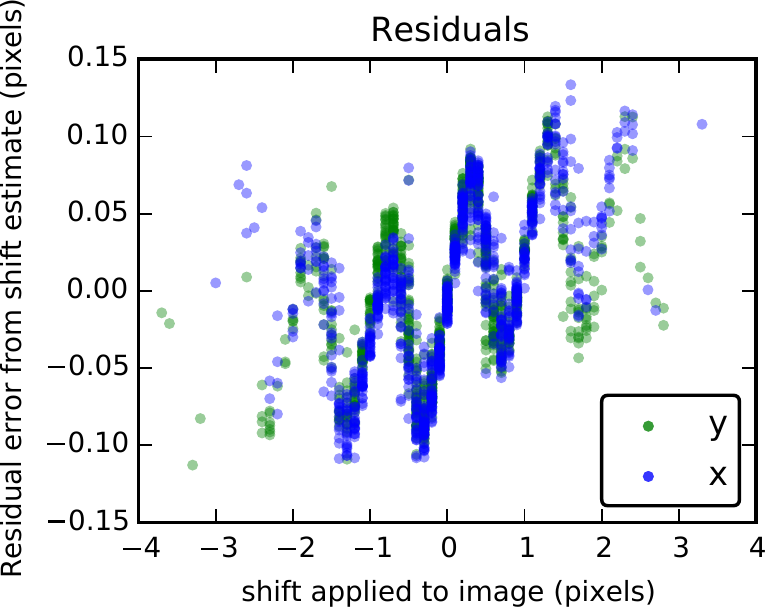}
                   }
    \caption{Measured image shift plotted against the actual shift applied to images. The negative y shifts are plotted in \subref{fig:aliasing_a} to make them easier to distinguish. There is a ``wobble'' apparent in the two lines, which is more clearly visible as a systematic effect in \subref{fig:aliasing_b}, where the residuals are plotted and take a ``sawtooth'' like pattern. This aliasing effect arises from under-sampling the correlation peak.}
    \label{fig:aliasing}
\end{figure}

\begin{figure}
    \centering
        \subfigure[]{\label{fig:bar_alias_a}
                     \includegraphics{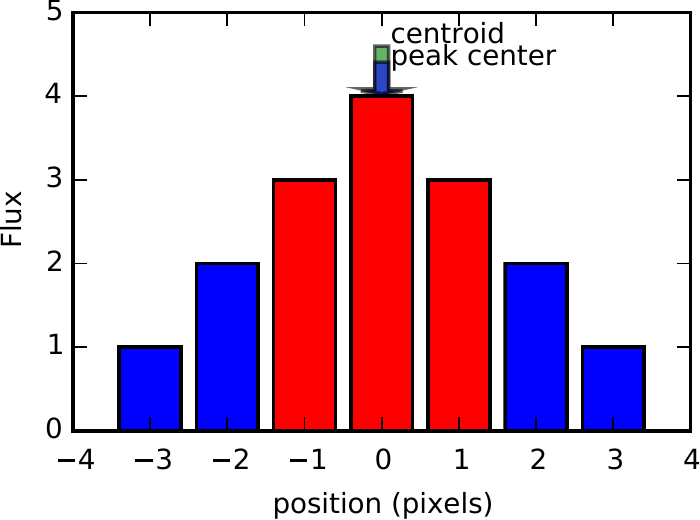}
                     }
        \\
        \subfigure[]{\label{fig:bar_alias_b}
                     \includegraphics{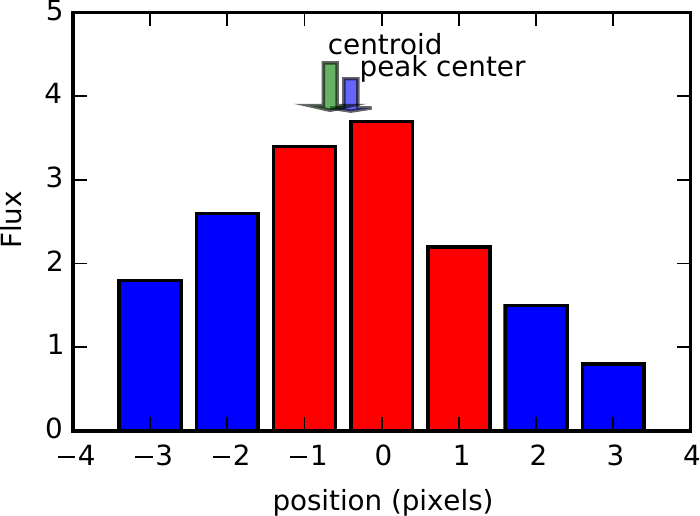}
                     }
    \caption{An illustrative 1D cut through a correlation peak, with the region used by the parabolic fit highlighted in red. Using only 3 pixels around the correlation peak, the shift estimate can be unavoidably biased away from the true location of the peak. While \subref{fig:bar_alias_a} shows the ideal case for using this method, there are some cases where the shift differs from the measured position due to the limited size of the region used, as demonstrated in \subref{fig:bar_alias_b}. This is shown by the arrows above the plots, the green arrow indicates where the parabolic centroid estimates the correlation peak, while the blue arrow shows the true location of the peak.}
    \label{fig:bar_alias}
\end{figure}

In high \ac{SNR} the limiting error in this technique arises from the biased sampling of the core of the correlation peak, illustrated in Fig.~\ref{fig:aliasing}. The sampling of the correlation peak results in a systematic rounding effect which biases the shift estimates towards integer values. The cause of this error is apparent in Fig.~\ref{fig:bar_alias}. Here we see the regions windowed for use in the centroid highlighted in red. This is a good mask for Fig.~\ref{fig:bar_alias_a}, however centering on the brightest pixel in Fig.~\ref{fig:bar_alias_b} shows that the peak is being under sampled, and not taking into account the full shape of the peak, giving an incorrect estimate of the peak location.

\subsection{Windowed, adaptive thresholding Center of Mass}
\label{sec:adaptive}
The simplest way to avoid under sampling the correlation peak is to use a larger window, however this allows more noise into the shift estimate. The noise can be removed to some extent by using a threshold to reject contributions from parts of the signal, of a similar strength as the noise. For a given autocorrelation shape and noise level there will be an optimal window size and threshold value, which gives the best estimate of the image shift.

Our proposed method is a two step process. Initially a window is placed around the correlation peak, then a thresholded center of mass is taken of the windowed region. The size of the window function and the threshold value are variable for each set of images. The threshold value is taken as a fraction of the relative peak intensity (max-min of the whole correlation image). Re-normalising intensity for every image is sympathetic to the shape and size of the correlation peak, and ensures that proportionally the same amount of the core of the peak is used in every measurement of the image shift, reducing bias effects.

The correlation image initially has a threshold applied, where pixels are rejected if their intensity falls below the threshold level, defined by:
\begin{equation}
    I_{\mathit{thresh}} < \left(I_{\mathit{max}} - I_{\mathit{min}}\right) \times pct,
    \label{eqn:threshold}
\end{equation}
where $I_{\mathit{thresh}}$ is the threshold intensity, $I_{\mathit{max}}$ is the maximum intensity in the correlation image, $I_{\mathit{min}}$ is the minimum intensity of the correlation image and $pct$ is the fractional threshold value. The thresholded correlation image then is masked to the chosen window size and is background subtracted, where the background value is the threshold intensity. The centroid estimate of image $i$, using a reference image $r$, can be described as a vector $\mathbf{R}^{i, r}$:
\begin{equation}
    \mathbf{R}^{i, r} = \left[
        \begin{array}{l}
            x_0 \\
            y_0
        \end{array}
    \right]
\end{equation}
where $x_0$ and $y_0$ are the $x$ and $y$ components of the centroid estimate $\mathbf{R}^{i, r}$. $\mathbf{R}^{i, r}$ is calculated using:
\begin{equation}
    \mathbf{R}^{i, r} = \frac{1}{I} \sum^{y_{\mathrm{max}}}_{y=1} \sum^{x_{\mathrm{max}}}_{x=1} I_{x,y} \mathbf{R}^{i, r}_{x,y}
    \label{eqn:com}
\end{equation}
where $I$ is the total intensity of the correlation image, $I_{x,y}$ is the intensity of pixel $x, y$ in the correlation image with the threshold applied, and $\mathbf{R}^{i,r}_{x,y}$ is the vector position of $[x, y]$ in the correlation image.

The size of the window is a relatively small parameter space to explore, going from a single pixel around the core (corresponding to an integer shift measurement), to the wings of the correlation peak succumbing to background noise. If any larger boxes are used a drop off in performance is seen as more noise is included in the centroid estimate, without any extra useful information being added. The outer threshold for this parameter needs to be set arbitrarily. If too small a window is used, a similar effect to the windowed parabolic fit is seen, in that the measurements are biased toward integer shifts. The optimum window size is chosen as a trade off between including as much of the correlation peak as possible, but also minimizing the number of pixels which only contribute noise to the measurement.

The centroid threshold value is normalised such that a value of 1 uses only pixels with the maximum flux and a threshold of 0 uses all available pixels. This parameter behaves similarly to the window size, in that using more pixels increases the noise contribution, reducing the accuracy of the shift estimate. Using high thresholds gives rise to a bias towards integer shift measurements, similar to that seen in the parabolic fit. The optimum threshold value lies somewhere between these two regimes, and is liable to change depending on the window size. This means the whole parameter space needs to be explored for all window sizes to identify the best combination of parameters for the centroids. We optimise the threshold and window size for a set of images, which all use a common reference image. The parameters then only need to be updated when the reference image is changed to take into account slow changing effects, such as the evolving granulation pattern on the solar surface.

The optimum set of parameters will depend on a number of different obvious factors, including the image shape, the shape of the resulting correlation function, and the \ac{SNR} of the images, assuming an arbitrary unknown correlation shape. There is no obvious analytic way to determine the best parameters for a given set of images, or circumstances, hence we explore the parameter space to find the optimal solution. However once the optimum set of parameters is found for a given object, at a set \ac{SNR} level, then it should be constant until one of these factors changes. In solar \ac{AO} the regions used for wavefront sensing are constantly evolving, causing the reference image used to be updated on a frame by frame basis. This also means that over time the optimum parameters are subject to change and need to be updated. As the parameters chosen are based on normalised intensity, they are insensitive to changes in flux for a given \ac{SNR}, such as scintillation effects.

\subsection{Error estimation}
\label{sec:estimation}
Given a set of shifted images of matching content and \ac{SNR}, it is possible to make multiple independent estimates of the image shift. By comparing the spread of the shift estimates we can get an estimate of the error on the shift measurement. Using different reference images allows us to estimate the shift in the image multiple times, we can then use the standard deviation of the shift estimates as an indicator of the error on the shift estimate. As this is a statistical process, the estimated error will not be accurate for the shift estimate of a single image in the set. However when averaged over the set of images, we can estimate the magnitude of the shift error on the set.

This set of images may be drawn from a single temporal wave-front sensor frame in \ac{AO}, guaranteeing spatial similarity of the images. Alternatively the set could be drawn from a time sequence in correlation video tracking. Care must be taken that the object does not change its spatial characteristics significantly over the duration of the set.

We use multiple different reference images, \emph{e.g.} using the first 10 sub-aperture images in the wave-front sensor frame, and use each of them as a reference to estimate image shifts. The global tip/tilt terms are then removed to compensate for the systematic error in shift estimation, due to the unknown shift applied to the reference image. This is a common practice in \ac{AO} systems to negate effects like wind shake from measurements. The subtraction of the global tip/tilt term can be described with:
\begin{equation}
    \mathbf{R}^{r}_{\mathit{t/t}} = \mathbf{R}^{r} - \left<{\mathbf{R}^{r}}\right>_{r},
    \label{eqn:average}
\end{equation}
for a given reference image, where $\mathbf{R}^{r}_{\mathit{t/t}}$ is the center of mass estimate of a set of images with tip/tilt removed, and $\left<{\mathbf{R}^{r}}\right>_{r}$ is the averaged tip/tilt term over all of the images using a given reference. This removes the shift due to each of the reference images, making the centroid estimates from different reference images directly comparable. The standard deviation of the resultant shifts estimate the error, $\sigma_{\mathbf{R}^{r}_{\mathit{t/t}}}$. This method of estimating centroiding errors allows for the parameter space to be explored on real data, where the actual shifts are unknown, and not just on simulated data.

\section{Results}
\label{sec:results}
\begin{figure}
    \centering
        \subfigure[]{\label{fig:paramspace_a}
                     \includegraphics{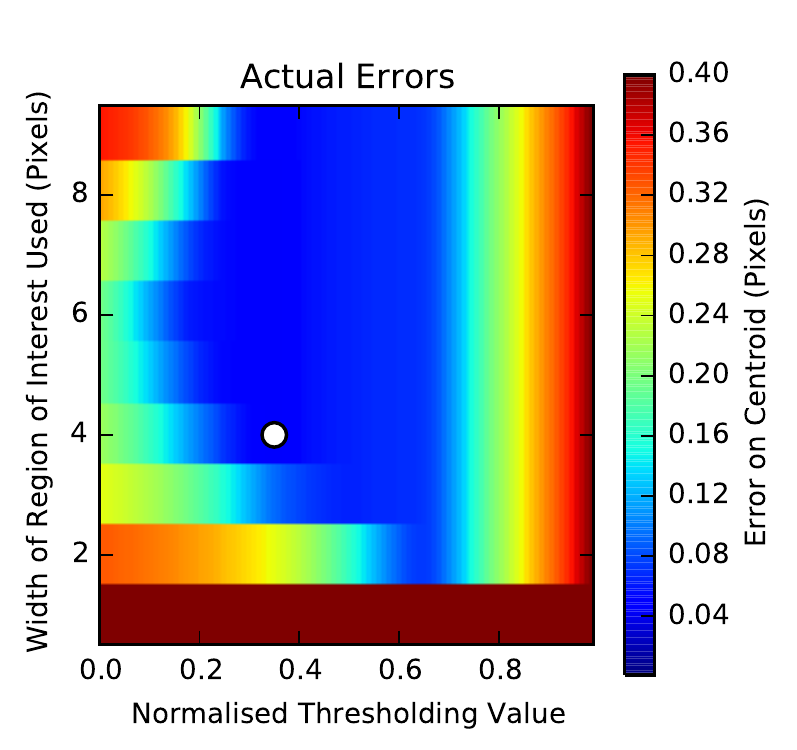}
                     }
        \\
        \subfigure[]{\label{fig:paramspace_b}
                     \includegraphics{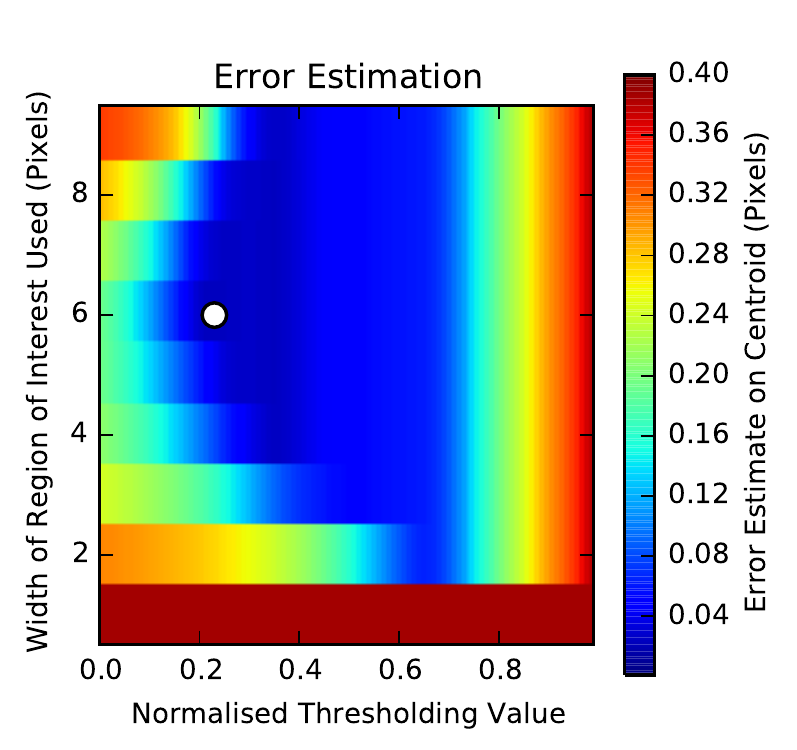}
                     }
    \caption{Full parameter space for the box size and threshold value in the center of mass algorithm. \subref{fig:paramspace_a} shows the real error associated with the parameters used in the center of mass technique, and \subref{fig:paramspace_b} gives the error estimate taken from the standard deviation on centroids using multiple reference images. The shape of the two plots is similar, indicating multiple references is a suitable estimator of the error. The white spots on the plot show where the optimum parameters lie for the respective methods. The estimated error position does not directly overlap with the location of the real minima, but it can be seen that the difference in error is minimal.}
    \label{fig:paramspace}
\end{figure}

The full parameter space was explored in simulation for a range of threshold values and window sizes applied to the correlation images. Fig.~\ref{fig:paramspace_a} shows the magnitude of residual errors for different sets of parameters. Fig. \ref{fig:paramspace_b} shows the standard deviation of the centroid measurements using ten different reference images. This has the same characteristics as the real error values, showing it can be used to estimate the location of the optimum parameters for the centroiding algorithm. The optimal parameters from each of the methods is highlighted with a white marker.

In the thresholding axis ($x$) of Fig.~\ref{fig:paramspace} it is possible to see the effects of aliasing towards the large thresholding values on the right of the plots. This effect is similar to the aliasing in the parabolic fit, and in all cases the error approaches that of integer pixel estimation, as at the largest threshold value only the brightest pixel is considered, equivalent to an integer pixel shift estimation.

In the window size axis ($y$) of Fig.~\ref{fig:paramspace} the structure is more complicated. Initially the aliasing is apparent for small window sizes, similar to the parabolic fit. This problem decreases as the window size increases, until its optimal region. However the performance begins to degrade again for large windows for low thresholding values. This happens where the region is so large that as well as including all of the peak of the correlation, it includes increasing amount of noise, which isn't filtered out by the thresholding.

\begin{figure}
    \centering
        \subfigure[]{\label{fig:SNR_detail_a}
                     \includegraphics{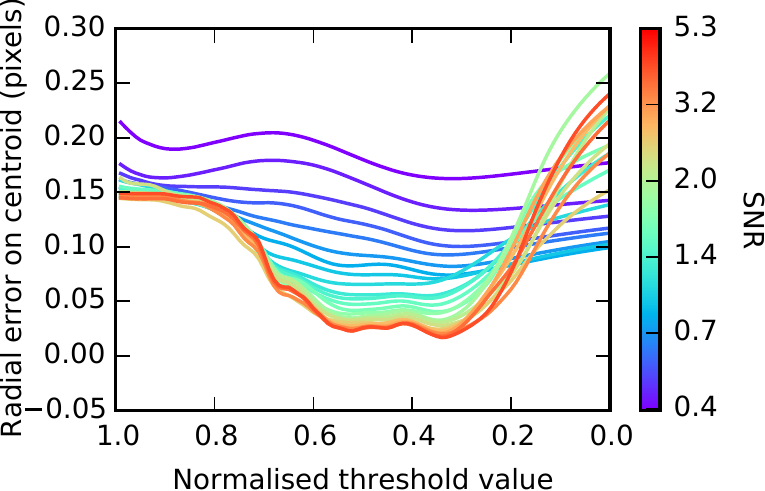}
                     }
        \\
        \subfigure[]{\label{fig:SNR_detail_b}
                     \includegraphics{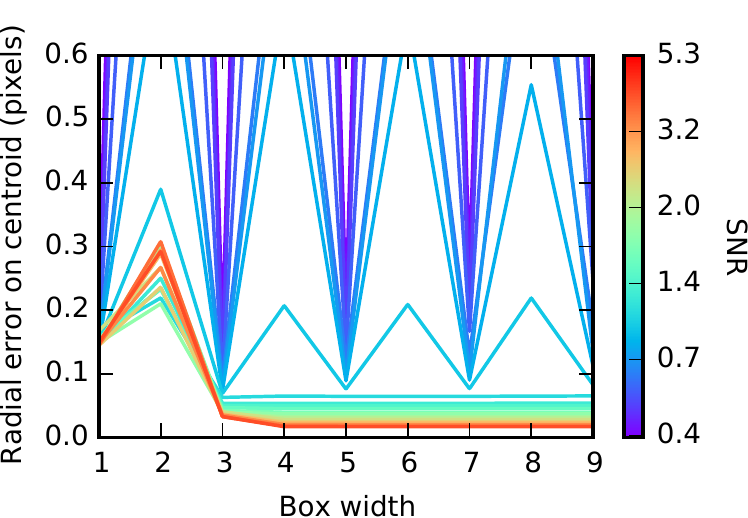}
                     }
    \caption{\subref{fig:SNR_detail_a} shows how the optimum threshold value is affected by different \ac{SNR} levels. \subref{fig:SNR_detail_b} shows how the window size  affects the error on the centroid estimate for different \ac{SNR} levels. Above a \ac{SNR} of 5 the curves no longer change, staying at their high SNR shapes.}
    \label{fig:SNR_detail}
\end{figure}

The centroid optimisation was performed on a range of different noise levels (using photon-noise) to demonstrate how noise affects the centroid estimates. The parameters dependence on \ac{SNR} is demonstrated in Fig.~\ref{fig:SNR_detail}, with Fig. \ref{fig:SNR_detail_a} showing how the threshold level affects the accuracy of the centroid estimates, and Fig. \ref{fig:SNR_detail_b} illustrating how changing the window size affects the accuracy of the centroid estimates. The estimation was performed on ten different regions of the granule image, with the errors taken to be the standard error.

\begin{figure}
    \centering
        \subfigure[]{\label{fig:best_params_a}
                     \includegraphics{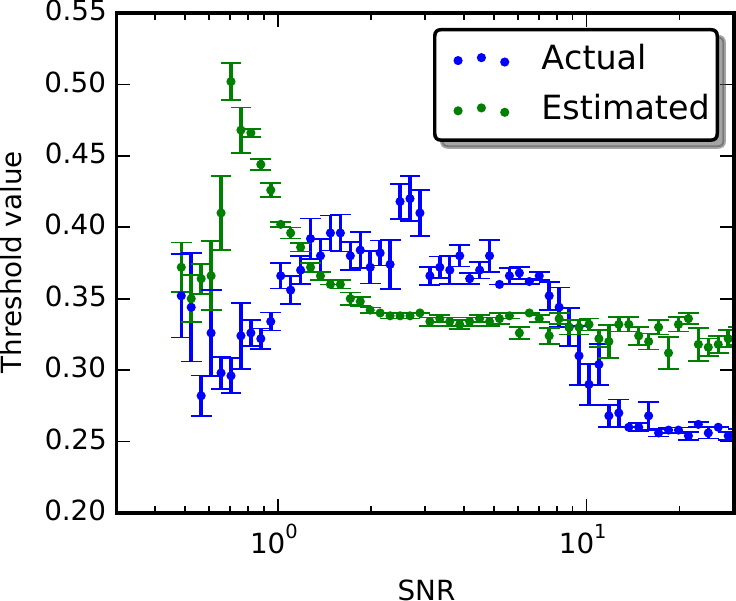}
                     }
        \\
        \subfigure[]{\label{fig:best_params_b}
                     \includegraphics{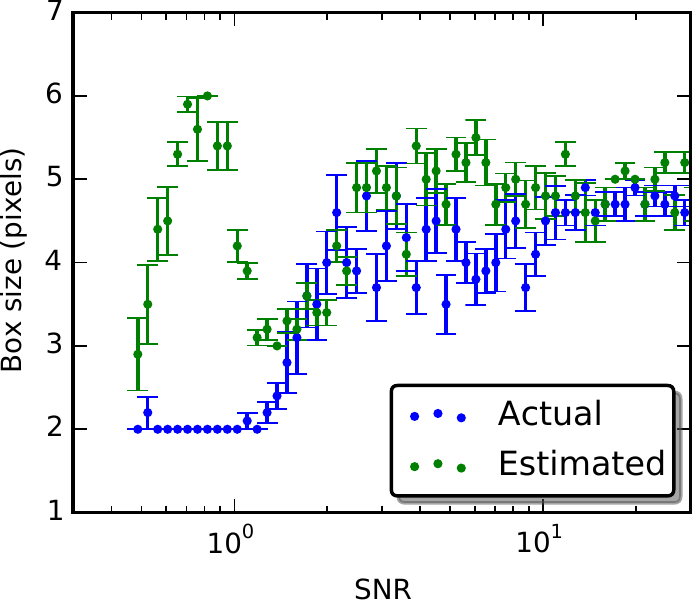}
                     }
    \caption{\subref{fig:best_params_a} shows the optimal thresholding value for the different \ac{SNR}s of the images used in the centroiding. Initially the thresholding is high, to remove as much noise as possible from the correlation image, then the thresholding drops to its optimum value for images which have low noise. \subref{fig:best_params_b} shows the box size for the different \ac{SNR} levels. This shows a similar tend, of increasing window size at high \ac{SNR}, using more pixels when the noise is reduced. At low \ac{SNR} the estimated parameters disagree with the true optimal parameters, but this disagreement decreases at higher \ac{SNR}.}
    \label{fig:best_params}
\end{figure}

The optimal values for the parameters varied with  \ac{SNR} as can be seen in Fig.~\ref{fig:best_params}. Fig. \ref{fig:best_params_a} shows the optimal thresholding values for the various \ac{SNR} levels, both best performing and the best estimated. The estimated threshold levels differ from the true value up to a SNR of 2, where the estimated threshold value is consistent, and in a region where small variations have little effect on the accuracy of the shift estimate. This trend is also seen in Fig. \ref{fig:best_params_b}, at low \ac{SNR} levels the estimated box size is larger than the actual optimal value, but at higher \ac{SNR} levels they agree more.

The optimal parameters for thresholding and window generally reduce the number of pixels used in the centroid in low \ac{SNR} conditions by using small thresholds and small window sizes to reduce the amount of noise in the centroid. At higher \ac{SNR} values the parameters stabilise for a given set of images to give the most accurate shift estimate.


Our technique fails in the low \ac{SNR} regime. This is due to the different sources of noise in the correlation image, and our sampling of them. The simplest way to do this is to have the images and their noise terms separate, as in equation~\ref{eqn:noise}:
\begin{equation}
    \begin{array}{ll}
        \mathrm{Im} &= \mathrm{Im}_{signal} + \mathrm{Im}_{noise} \\
        \mathrm{Ref} &= \mathrm{Ref}_{signal} + \mathrm{Ref}_{noise}
    \end{array}
    \label{eqn:noise}
\end{equation}
where $\mathrm{Im}$ represents the overall image being centroided, $\mathrm{Im}_{signal}$ describes the signal in the image, and $\mathrm{Im}_{noise}$ describes the noise associated with the image, in our case shot noise. $\mathrm{Ref}$ follows similar definitions for the reference image. When combined, assuming a linear regime, the correlation image has four terms:
\begin{multline}
    \mathrm{Corr} = \mathrm{Corr}\left[_{\mathbf{Im}_{signal} \mathbf{Ref}_{signal}}\right] + \mathrm{Corr}\left[_{\mathbf{Im}_{signal} \mathbf{Ref}_{noise}}\right] \\
    + \mathrm{Corr}\left[_{\mathbf{Im}_{noise} \mathbf{Ref}_{signal}}\right] + \mathrm{Corr}\left[_{\mathbf{Im}_{noise} \mathbf{Ref}_{noise}}\right]
\end{multline}
where $\mathrm{Corr}$ is the total signal in the correlation image, with the contributing factors all described to the right. If we assume that the contribution of $\mathrm{Corr}\left[_{\mathbf{Im}_{noise} \mathbf{Ref}_{noise}}\right]$ is negligible, then there are two remaining error terms which affect our estimate of the centroid. However by taking an average over different references in our estimate of the error, we are in effect averaging out the $\mathrm{Corr}\left[_{\mathbf{Im}_{signal} \mathbf{Ref}_{noise}}\right]$ term. This term becomes more dominant at lower \ac{SNR} levels, hindering the performance of our technique. There are other methods of estimating the error of a centroid on an extended object, such as \citet{Saunter2010}, which don't have this problem, but this requires an oversampling of the correlation peak, something avoided in \ac{AO} to reduce data rates and computation time.

\begin{figure}
    \begin{center}
        \includegraphics{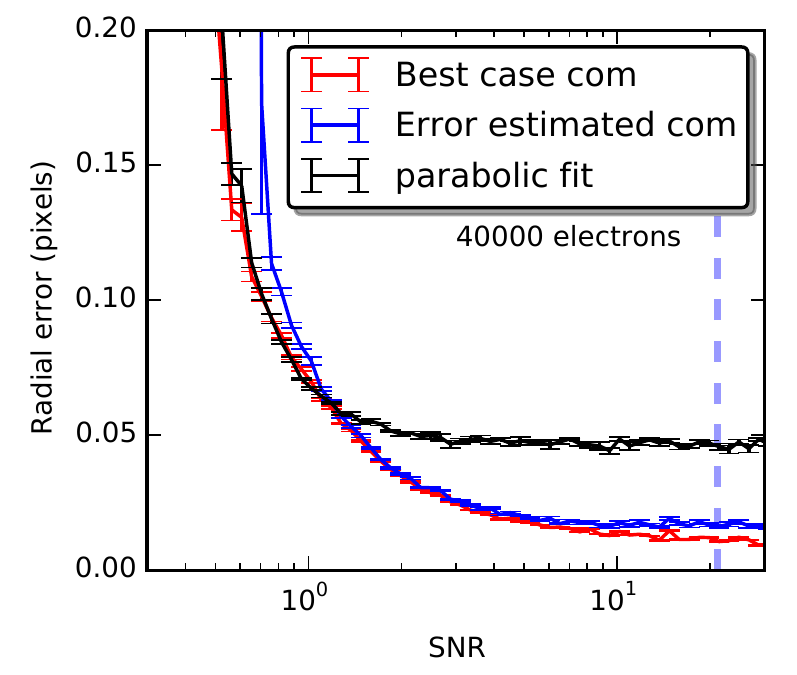}
    \end{center}
    \caption{This plot shows the performance of the center of mass algorithms and the 2D parabolic fit for a range of different \ac{SNR}s. It can be seen above a \ac{SNR} of 1 the windowed, thresholded center of mass outperfroms a 2D parabolic fit. The 2D parabolic fit tapers off in performance at 0.05 pixel error, whereas the windowed center of mass has a much lower performance threshold. The vertical line on the plot show the expected \ac{SNR} for a solar granule image with a contrast of 10\%, and a camera with a full well depth of 40000 electrons, which represents typical conditions in solar \ac{AO}. It can also be seen that the performance from estimating the errors on the center of mass is worse than the optimal case, but does still reach close to peak performance.}
    \label{fig:SNRs}
\end{figure}

The overall performance of the centoriding techniques for the different \ac{SNR}s is shown in Fig.~\ref{fig:SNRs}. For high \ac{SNR}, the best performance is given by the thresholded, windowed center of mass measurements, with little difference between the theoretical best performance and the performance derived from error estimation. The overall boost in accuracy is $3 \times$ for the high \ac{SNR}. For \ac{SNR} below 1, the windowed parabolic fit outperforms the thresholded, windowed center of mass method. This could be due to the crude error estimator implemented here and it may be possible to improve this using other error estimation techniques \citep{Saunter2010}. However this still could only bring the performance back to the level of the 2D parabolic fit at best. Our technique is best suited to high \ac{SNR} regimes.

\section{Conclusions}
\label{sec:conclusions}
We have demonstrated that for tracking extended sources, a method of error estimation allows different centroiding parameters to be explored on real data, allowing for the optimum parameters to be chosen. While this does take extra computation, the correlation images only needs to be generated once for each reference, minimizing the increase in computation effort required. Also once the optimum set of parameters has been found, they should hold as the best parameters until something in the system changes, i.e. a change of target, or reference image. The parameters for the centroiding algorithm should be updated regularly to keep it optimal.

Exploring the parameter space is a parallelisable process, so can be performed quickly. With the use of SIMD \citep{furht2008} and more advanced optimisation algorithms, rather than the brute force method exploring the full parameter space implemented here, the method should be viable for use in a real-time system.

The method of noise estimation used here is crude, though good enough for our purposes, and could be used for different parameters in other techniques, such as \citet{Li2008}. There are more efficient algorithms for estimating noise on centroiding of extended objects, such as \citet{Saunter2010}, which could also be implemented to give quantitative estimators of centroiding accuracy, as well as being computationally less intensive than the multiple reference approach.

Overall, for the solar case, with high \ac{SNR}, the use of an optimised, thresholded, windowed center of mass algorithm offers a factor of $3 \times$ improvement in centroiding accuracy over the windowed parabolic fit. This could be used real-time in solar \ac{AO} for better wave-front estimations, and also with post processing techniques, such as measuring more accurate atmospheric profiles.

Further investigation should be performed in the low \ac{SNR} regime, where both the center of mass and 2D parabolic fit methods give poor performance, to see if more accurate centroids can be extracted. There is also more work to be done in implementing the technique into a real system which performs centroiding on extended objects, to see how it affects system performance.

\section*{Acknowledgments}

M. J. T gratefully acknowledge support from the Science and Technology Facilities Council (STFC) in the form of a Ph.D studentship (ST/K501979/1). The authors would like to thank the Institute of Solar Physics, Sweden, Mats Carlsson, Viggo Hansteen, Luc Rouppe van der Voort, Astrid Fossum, and Elin Marthinussen for taking the raw image used in this paper, and Mats L\"{o}fdahl, for performing the image reconstruction to produce the final image. M. J. T. would like to thank Prof. Gordon Love for all of his advice and guidance, and the referee for their insightful advice and feedback. Data used is available from the author on request.

\bibliographystyle{mn2e}
\bibliography{paper}

\label{lastpage}

\end{document}